# Graphical Join: A New Physical Join Algorithm for RDBMSs

Ali Mohammadi Shanghooshabad, Peter Triantafillou
University of Warwick
{ali.mohammadi-shanghooshabad,p.triantafillou}@warwick.ac.uk

## ABSTRACT

Join operations (especially n-way, many-to-many joins) are known to be time- and resource-consuming. At large scales, with respect to table and join-result sizes, current state of the art approaches (including both binary-join plans which use Nested-loop/Hash/Sort-merge Join algorithms or, alternatively, worst-case optimal join algorithms (WOJAs)), may even fail to produce any answer given reasonable resource and time constraints. In this work, we introduce a new approach for n-way equi-join processing, the Graphical Join (GJ). The key idea is two-fold: First, to map the physical join computation problem to PGMs and introduce tweaked inference algorithms which can compute a Run-Length Encoding (RLE) based join-result summary, entailing all statistics necessary to materialize the join result. Second, and most importantly, to show that a join algorithm, like GJ, which produces the above join-result summary and then desummarizes it, can introduce large performance benefits in time and space. Comprehensive experimentation is undertaken with join queries from the JOB, TPCDS, and lastFM datasets, comparing GJ against PostgresQL and MonetDB and a state of the art WOJA implemented within the Umbra system. The results for in-memory join computation show performance improvements up to 64×, 388×, and 6× faster than PostgreSQL, MonetDB and Umbra, respectively. For on-disk join computation, GJ is faster than PostgreSQL, MonetDB and Umbra by up to 820×, 717× and 165×, respectively. Furthermore, GJ space needs are up to 21,488×, 38,333×, and 78,750× smaller than PostgresQL, MonetDB, and Umbra, respectively.

## 1 INTRODUCTION

Join queries are essential in Relational Data Base Systems (RDBMSs). However, they are known to be expensive operations especially for multi-way joins with many-to-many relationships (involving joins on non-key attributes) which are frequent in analytics. In these cases, the state of the art approaches suffer in both time and space. The main source of inefficiency stems from *algorithms exerting redundant effort*. First, some produce intermediate join results (say, after binary joins) which contain tuples which do not make it in the final n-way join result. Second, even the final join result typically has a lot of redundancy which introduces space and time inefficiencies.

One can distinguish two use cases for physical join algorithms. In "compute-and-forget", the join result is to be computed once, typically in memory (and then likely forgotten). In "compute-and-reuse", a join result is to be stored on disk (e.g., in order to be re-used in the future). In both cases generating unnecessary intermediate tuples and/or computing, storing, and retrieving redundant tuples incurs significant time-space overheads.

Table 1

| ... | A | B | ... |
|---|---|---|---|
| ... | $a_0$ | $b_0$ | ... |
| ... | $a_0$ | $b_0$ | ... |
| ... | $a_0$ | $b_0$ | ... |
| ... | $a_1$ | $b_1$ | ... |
| ... | $a_1$ | $b_1$ | ... |
| ... | $a_2$ | $b_1$ | ... |
| ... | $a_3$ | $b_3$ | ... |
| ... | $a_3$ | $b_3$ | ... |
| ... | $a_3$ | $b_4$ | ... |
| ... | $a_3$ | $b_4$ | ... |
| ... | $a_3$ | $b_4$ | ... |
| ... | $a_3$ | $b_4$ | ... |

Table 2

| ... | B | C | ... |
|---|---|---|---|
| ... | $b_0$ | $c_0$ | ... |
| ... | $b_0$ | $c_0$ | ... |
| ... | $b_1$ | $c_0$ | ... |
| ... | $b_1$ | $c_0$ | ... |
| ... | $b_1$ | $c_0$ | ... |
| ... | $b_2$ | $c_1$ | ... |
| ... | $b_2$ | $c_1$ | ... |
| ... | $b_2$ | $c_1$ | ... |
| ... | $b_3$ | $c_2$ | ... |
| ... | $b_4$ | $c_3$ | ... |
| ... | $b_4$ | $c_3$ | ... |
| ... | $b_4$ | $c_4$ | ... |

Table 3

| ... | C | D | ... |
|---|---|---|---|
| ... | $c_1$ | $d_0$ | ... |
| ... | $c_1$ | $d_0$ | ... |
| ... | $c_1$ | $d_0$ | ... |
| ... | $c_1$ | $d_0$ | ... |
| ... | $c_2$ | $d_2$ | ... |
| ... | $c_2$ | $d_2$ | ... |
| ... | $c_2$ | $d_2$ | ... |
| ... | $c_2$ | $d_2$ | ... |
| ... | $c_3$ | $d_3$ | ... |
| ... | $c_3$ | $d_3$ | ... |
| ... | $c_4$ | $d_4$ | ... |
| ... | $c_4$ | $d_4$ | ... |

**Figure 1: Three-table chain join**

Despite the large attention join algorithms have received, there is still lots to be done in order to extinguish all redundant effort in computing, storing, and retrieving join result tuples, which will minimize both time and space and will be applicable in both the compute-and-forget and compute-and-reuse scenarios. Specifically, the most popular physical join algorithms (such as the Nested-loop/Hash/Sort-merge Join algorithms and their derivatives [12–14, 20, 25, 30]) occupy a significant part of the code base in all RDBMS products. These algorithms become inefficient when handling multi-way joins with at least one many-to-many relationship in the joined tables. This inefficiency stems from their binary nature: they join two tables at a time [5, 53] and this produces intermediate join results which include tuples which not part of the final result. To address this problem, Ngo et al. [34, 35] introduced a new (worst case optimal) join algorithm which exactly avoids paying the cost to generate unneeded intermediate tuples. Nonetheless, even WOJA algorithms do not fully address the space concerns, as redundancy exists even in the final result. This is especially important in the compute-and-reuse case where extra space directly translate to extra time (to store and retrieve this redundant data).

Driven by the above considerations, this work will propose a new approach for n-way equi-joins with time and space gains and in both compute-and-forget and compute-and-reuse scenarios. Its fundamental idea is to first capture the key statistical information required to generate all join result tuples, but *without computing the actual expensive full join.* This effort will leverage PGMs and propose tweaked PGM inference algorithms which will generate a summary of the join result. This summary can be generated, stored to disk and loaded into memory faster than the actual join result, yielding significant gains for the compute-and-reuse scenarios [46]. Additionally, even for the compute-and-forget scenarios, this work will show that generating the summary and then desummarizing it outperforms all state of the art approaches (based on binary-join plans or WOJA).

### 1.1 The Problems

More formally, the two sources of inefficiency with physical join algorithms are as follows. Assume $T_1, T_2, T_3, ..., T_n$ are the tables

Full join result

| id | A | B | C | D |
|---|---|---|---|---|
| 0 | $a_3$ | $b_3$ | $c_2$ | $d_2$ |
| ... | ... | ... | ... | ... |
| 7 | $a_3$ | $b_3$ | $c_2$ | $d_2$ |
| 8 | $a_3$ | $b_4$ | $c_3$ | $d_3$ |
| ... | ... | ... | ... | ... |
| 23 | $a_3$ | $b_4$ | $c_3$ | $d_3$ |
| 24 | $a_3$ | $b_4$ | $c_4$ | $d_4$ |
| ... | ... | ... | ... | ... |
| 31 | $a_3$ | $b_4$ | $c_4$ | $d_4$ |

Summary of the join result

| A | B | C | D |
|---|---|---|---|
| $a_3$,32 | $b_3$,8 | $c_2$,8 | $d_2$,8 |
| | $b_4$,24 | $c_3$,16 | $d_3$,16 |
| | | $c_4$,8 | $d_4$,8 |

**Figure 2: Join result and GFJS**

to be joined. Given a query execution plan consisting of binary joins, call $S_{tmp} = \{R_{\{1,2\}}, R_{\{1,2,3\}}, \dots, R_{\{1,2,\dots,n-1\}}\}$ the set of all intermediate temporary tables. Any tuple $t \in R_t$ where $R_t \in S_{tmp}$ is called an intermediate tuple and if $t \in R_t$ but $t \notin R_{\{1,2,\dots,n\}}$, we call $t$ an Unneeded Intermediate Result (UIR). Generating UIR tuples is a key source of inefficiency. Especially when the size of any $R_t \in S_{tmp}$ becomes much larger than $R_{\{1,2,\dots,n\}}$.

Consider Figure 1: If Table 1 and Table 2 are joined first, there will be 15 tuples with $b_0$ and $b_1$ which will be ignored when they are joined with Table 3 because there is no $c_0$ in Table 3. This makes the binary join plans sub-optimal. Note that this sub-optimality still remains if the Table 2 and Table 3 are joined first because Table 1 does not include $b_2$, so this is not a join-ordering problem per se.

Another inefficiency stems from the redundancy in the join result itself (a by-product of denormalization). In Figure 2, column A has one distinct value ($a_3$), repeated 32 times, while $a_3$ is repeated only 6 times in (normalized) Table 1. Such redundancy leads to inefficient join algorithms (in time and space), especially in compute-and-reuse cases where results are to be stored and reloaded.

### 1.2 The Solution Approach

GJ aims to avoid these issues based on the following summary.

*Definition* 1 (Grouped Frequentist Join Summary (GFJS)). GFJS is a result of the summarization over the grouped join result per column. It is formed by replacing the repeated values, $v$, with pairs $(v, freq)$ where $v$ is the value and $freq$ is its frequency.

One may view GFJS as follows. Take the join result $R_{\{1,2,\dots,n\}}$ with columns $A = \{A_0, A_1, \dots, A_m\}$ and then sort it according to all columns in $A$. Then, for each column produce, an RLE encoding of its values: starting from the top of each $A_i \in A$, replace all repeated values $v$ in the column with the pair $(v, freq)$. This will yield GFJS.

Figure 2 shows GFJS for the join of the three tables in Figure 1. The join result in that figure is already sorted, so per column in the join result we can start from the top and replace the repeated values with a pair of the value and its frequency. For column $A$, there is one distinct value ($a_3$) with frequency of 32. For column $B$, there are two distinct values, $b_3$ and $b_4$, with the frequencies of 8 and 24, respectively. The sum of all frequencies in different columns should, of course, be equal.

The goals are to: (i) generate GFJS without computing the (expensive) join – for this we will leverage PGMS; (ii) produce the join result from GFJS. The latter entails an optional step to store/retrieve GFJS and then desummarize it.

**Why PGMs?** Calculating GFJS without calculating the join result first is essentially like calculating the marginals for attributes over the normalized tables. In Figure 2, $(a_3, 32)$ in GFJS is just the marginal for A over the join result. PGMs by construction facilitate the global understanding (in our case, the marginals) from local observations (in our case, normalized tables) in an efficient way. Obviously, extracting GFJS from a distribution is easier and more understandable than extracting it from raw data because with distributions we deal with probabilities (or, equivalently frequencies) and GFJS is all about frequencies. So, the only thing needed is to learn the distribution within single tables and rely on PGMs to calculate the marginals over the join result. And, learning distributions from tables (local observations) is easily/efficiently done by scanning the tables once and calculating the exact frequencies for attributes. More formally, assume the attribute order in generating GFJS is $\{A_0, A_1, \dots, A_m\}$. To generate GFJS, we need $p(A_0)$, $p(A_1|A_0)$, ..., $p(A_m|A_{m-1})$ over the joint distribution of the to-be-joined tables. These probabilities can be extracted from the marginals of paired attributes and PGMs can efficiently calculate those marginals.

### 1.3 Contributions

This work introduces:

- A different way of thinking about physical n-way equi-join algorithms: Instead of building indexes and using binary-join algorithms or WOJAs over raw data tables, first produce a (PGM-based) join summary, followed by optionally storing it to disk and retrieving it from disk when called to (re)produce the join result, and finally desummarizing it.
- A novel algorithm to generate of an RLE-style summary over the grouped join result *without* executing the expensive join operation. (This summary has other applications of its own).
- A complexity analysis showing worst-case optimality.
- A detailed performance evaluation using JOB, lastFM and TPCH data and queries, comparing GJ against join processing in PostgreSQL [15], MonetDB [23]) and the WOJA in Umbra [17], showcasing the new approach's large gains in space and time.

The code for GJ is available at https://github.com/shanghoosh1/Graphical_join.

In what follows, Section 2 presents the key background for worst-case optimal join algorithms and PGMs. Section 3 presents the new concepts, structures, and algorithms in GJ, which extend and complement those in the background. Section 4 studies experimentally the performance of GJ against three RDBMS baselines. Sections 5 and 6 discuss related work and conclude the paper.

## 2 BACKGROUND

### 2.1 Worst-Case Optimal Join Algorithms

WOJAs aim to deal with UIR (but not with redundancy in the final join result) in n-way joins with many-to-many relationships. Any join algorithm with a complexity of $O(N^{\rho})$ is a WOJA, where $\rho$ is the fractional edge cover.

2.1.1 *The Fractional Edge Cover $\rho$.* Consider $N$, the size of the largest table in a join. Assume a hyper-graph $H(V,E)$ is constructed over the involved tables; $V$ is the set of all nodes (a node per attribute) and $E$ is the set of the hyper edges among them: For each table, there is an hyper-edge connecting all the attributes coming from the same table. The edge cover is a subset $E_C \subseteq E$ of edges such that each attributes appears in at least one edge. Finding the fractional edge cover $\rho$ can be formulated as a linear programming problem by assigning to each edge $e_i \in E$ a weight $\lambda_i = [0-1]$. For example, for a "triangle" join query $TQ$ of tables $T_1(A1,A2), T_2(A2,A3), T_3(A3,A1)$ with the sizes $|T_1|, |T_2|$ and $|T_3|$ respectively, the linear program can be defined as below:

Minimize $\lambda_{T_1} + \lambda_{T_2} + \lambda_{T_3}$, subject to

$$A1: \lambda_{T_1} + \lambda_{T_3} \geqslant 1$$
$$A2: \lambda_{T_1} + \lambda_{T_2} \geqslant 1$$
$$A3: \lambda_{T_2} + \lambda_{T_3} \geqslant 1$$

For example, choosing $\lambda_{T_1} = \lambda_{T_2} = \lambda_{T_3} = 1/2$ for $TQ$ yields a valid fractional edge cover and the upper bound is $O(N^{3/2})$. [21] showed that the product of each table size to the power of $\lambda$s is larger or equal to $|Q|$ and at the same time it is a tight upper bound. For $TQ$:

$$|T_1|^{\lambda_{T_1}} \times |T_2|^{\lambda_{T_2}} \times |T_3|^{\lambda_{T_3}} \geqslant |Q| \quad (1)$$

WOJAs have enjoyed significant attention in the last decade and several WOJAs have been introduced e.g. NPRR in [34], Leapfrog TrieJoin in [49] and the versions of WOJAs with new data structures to build the tries including [17], [4], [16]. A general pseudo-code for all WOJAs can be found in Algorithm 1 in [17], which offers also a state-of-the-art WOJA for RDBMS.

## 2.2 Probabilistic Graphical Models (PGMs)

At its heart, SGJ relies on probabilistic inference. Probabilistic inference is essentially a task of calculating quantities (e.g. marginals) of some variables over a distribution [26].

More specifically, a PGM represents a distribution in a factorized way with a graph $G(V,E)$ and some rules $\mathcal{M}$, where $V$ is a set of vertices (a.k.a. variables or nodes) and $E$ is a set of edges among the nodes. For example, a Markov Random Field (MRF) is a type of PGM having an undirected graph $G$ and three (Pairwise, Local and Global) Markov properties as $\mathcal{M}$ which are defined by the concept of *conditional independence.* Two variables $A$ and $B$ are conditionally independent given $C$ if $P(A,B|C) = P(A|C) \times P(B|C)$.

*Definition* 2 (Cliques C(G)). A clique c is a set of fully connected nodes in a graph G(V,E). C(G) contains all the cliques that exist in G. In other words, if $c \in C(G)$, and $u, v \in c$ then the edge $(u,v) \in E(G)$

Given a graph $G(V,E)$ with cliques $C(G)$, the probability distribution $p(x_V)$ of an MRF can be represented as follows

$$p(x_V) = \frac{1}{Z} \prod_{c \in C(G)} \phi_c(x_c) \quad (2)$$

where $\Phi$ is a factor (a.k.a. potential function), which is a non-negative function over the joint domain of a set of variables. $p(x_V)$ is the product of all factors of cliques in the MRF. $Z$ is called the partition function which is a normalization constant:

$$Z = \sum_{x_V} \prod_{c \in C(G)} \phi_c(x_c) \quad (3)$$

In our case, $Z$ is the join size. Since we just need the frequencies of distinct values (not the probabilities) to generate the join summary/result, we can omit $Z$ from the Equation 2.

*Definition* 3 (Maximal Cliques/ maxcliques $\mathcal{C}(G)$). A maximal clique (or maxclique) is a clique which by adding any $v \in V$ in that clique, makes it no longer a clique.

Any clique can be absorbed by its maxclique. So, we can replace $C$ with $\mathcal{C}$ in Equation 2. That absorption is accomplished by calculating the product of the cliques inside a maxclique.

Not only PGMs can reduce the storage cost by keeping smaller local potentials, but also they make the probabilistic inference (e.g. marginalization) more efficient by algorithms such as Variable Elimination Algorithm (VEA). VEA is an inference algorithm based on dynamic programming, and it is employed to calculate the marginals over the factorized distribution represented by a PGM.

**VEA:**To calculate a marginal over the distribution presented in Equation 2, VEA eliminates the variables one by one based on an elimination order $O$ by using a sum-product operation presented in Equation 4 per variable. For example, lets assume we have $n$ variables in $V = \{x_1, x_2, ..., x_n\}$. To calculate the marginal of a variable $x_1$, all other variables should be eliminated (summed-out) one-by-one. The formula for the sum-product operation for eliminating one of them $(x_i)$ is as follows:

$$\sum_{x_2,\ldots,x_{i-1},x_{i+1},\ldots,x_n} \prod_{c \in O_{x_i}} \phi_c(x_c) \sum_{x_i} \prod_{c \in \mathcal{I}_{x_i}} \phi_c(x_c) \quad (4)$$

where $\mathcal{I}_{x_i}$ and $O_{x_i}$ contain the cliques that do and do not involve $x_i$, and $\phi$ is the potential function.

Thus, we can say that the complexity of calculating the marginals is specified by the size of the largest $\mathcal{I}$ (how many variables it contains). Assume $\mathcal{I}$ has $m$ variables and each variable has $r$ distinct items in its domain, the complexity of the VEA to eliminate all variables is $O(r^m)$.

Note, that after eliminating a variable, all its neighbors should make a single clique, and if the neighbors are not fully-connected in the corresponding graph, new edges are added. The new edges are called fill-in edges. These new edges makes the inference inefficient and we want to avoid them because they make larger cliques and the larger cliques make larger $\mathcal{I}$ in the sum-product operation.

In trees, each node has just one parent, meaning that there is at least one elimination order $O$ that does not introduce any new fill-in edges. We call that $O$ a perfect elimination order. An elimination order is perfect when by eliminating the variables, no fill-in edges are added to the graph. Trees thus have a perfect elimination order starting from the leaves to the root (note, not all eliminations are perfect in trees), and VEA can calculate the marginals optimally. But what about on graphs (with cycles)?

2.2.1 *VEA on Graphs.* Generally, to run VEA on graphs, people translate the graph to a structure that has the same characteristics as trees. In other words, a tree of maxcliques is made from the graph, so that every node may have more than one variable. The Running Intersection Property (R.I.P) is one of the characteristics that the tree of maxcliques should hold, as it holds for trees. R.I.P is essential for SGJ in dealing with cyclic graphs.

**R.I.P of maxcliques:** Let $C_1, C_2, ..., C_l$ be an ordered sequence of maxcliques in graph G where $l$ is the number of maxcliques. We say the ordering obeys the R.I.P if for all $i$>1, there exists $j < i$ such that $C_i \cap (\cup_{k<i} C_k) = C_i \cap C_j$

In Figure 3 (e), assume the order is $C1, C2$ and $C3$ then the intersection between $C2$ and $C3$ (which is the node $(C, D)$) is equal to $C3 \cap (C2 \cup C1)$. The union $(C2 \cup C1)$ is called the history of $C3$.

**Junction Tree (JT) of maxcliques:** JT of maxcliques is a tree of the maxcliques which possesses the R.I.P property. There have exist many algorithms to translate graphs to JTs [10, 24, 26]. Here, we just provide a brief explanation of JT creation.

For a graph, there may be more than one JT. We need to find the best JT where the size of the largest maxclique is smaller than the size of largest maxcliques in other possible JTs from the same graph. Recall that the complexity of finding a marginal is dependent on the size of the largest clique in the graph. For example, we could take all the variables as one maxclique by adding fill-in edges, but in this case, the size of the largest maxclique would be equal to the size of $V$ and the complexity of the inference (VEA) on that tree will be exponentially high. Extracting the best JT is NP-Complete.

It has already been proved that if a graph is triangulated then it has a JT of maxcliques which possesses R.I.P. [6, 50]. In a triangulated graph, there should not be a cycle with more than 3 nodes. The triangulated graphs have a perfect elimination order and eliminating any node will not introduce any new fill-in edges.

Not all graphs have a perfect elimination order. For example, in Figure 3 (a), there is a cycle with 4 nodes and eliminating each of them will introduce a fill-in edge. If the graph is not triangulated (e.g. Figure 3 (a)), finding the best elimination order and adding new fill-in edges can lead us to the best triangulated graph, thus we can have the best JT as well. One should check all the combinations to find the best elimination order, hence finding the best elimination order is NP-Complete as well. The good news is that, in our join problem, the number of nodes is small and hence one can find the best elimination order with a small number of fill-in edges, manually. Nonetheless, greedy heuristic algorithms, such as min fill-in, work well even on the graphs with thousands of nodes.

**The Min fill-in heuristic:** In each step, the min fill-in heuristic adds one node in the elimination order $O$, and that node is the node which introduces the minimum number of new fill-in edges. If there are more than one node with minimum fill-in edges, it breaks the ties arbitrarily. In Figure 3, $A$ should be eliminated first then among $B, C, D$ and $E$, one is chosen randomly, and so on and so forth.

The min fill-in heuristic can provide the triangulated graph. For a given graph $G(V, E)$, we find the new fill-in edge set $E'$ with the min fill-in heuristic and make the triangulated graph as $G'(V, E \cup E')$. Figure 3 (b) is one of the possible triangulated graphs for the graph Figure 3 (a). Note that when we add a fill-in edge, it is like we add a potential equal to 1 that can join with any other potentials. The main potentials for fill-in edges are calculated during inference. The min fill-in heuristic can also output all the maxcliques in the graph during triangulation since after elimination of a node, all its neighbors should be a clique. Any pair of maxcliques that have some shared variables are connected to each other so that they make a graph of the maxcliques (e.g. Figure 3 (c)).

So, the output of the min fill-in heuristic contains an elimination order $O$, a triangulated graph, and a graph of the maxcliques. The issue is how to derive a JT of maxcliques from the graph of maxcliques. There are several ways, and the easiest way is to apply the maximal spanning tree algorithm. This algorithm finds all the separator sets among maxcliques and chooses the edges with maximum separator sizes one-by-one to span the graph of maxcliques. A separator set is a set of nodes in the graph $G$ that if we remove it, $G$ is divided into disconnected sub-graphs.

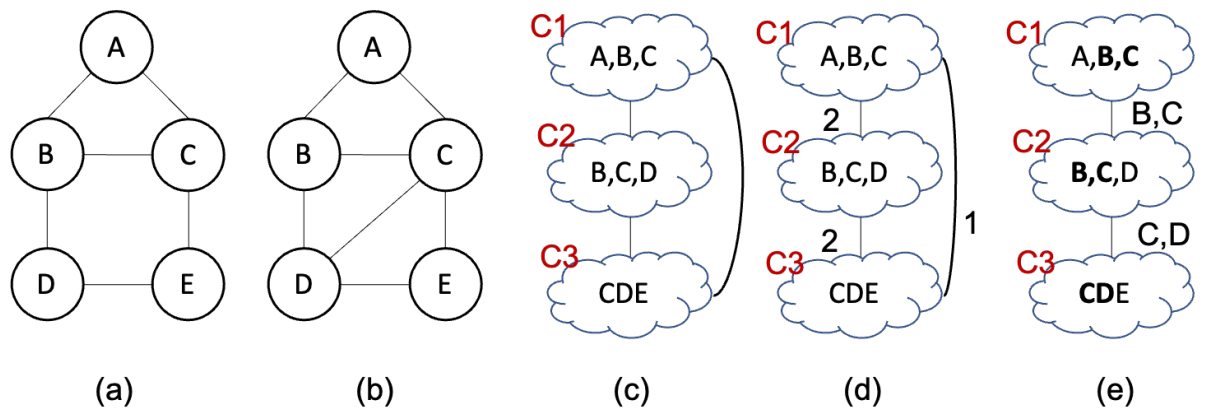

**Figure 3: Example non-tree graph translation to a tree**

Figure 3 shows all the steps of translating the graphs to JTs of maxcliques. (a) shows the original graph $G$, (b) shows the triangulated $G$, (c) contains the graph of maxcliques. (d) shows the weights (separator sizes) and finally (e) is the JT of maxcliques after applying the maximal spanning tree algorithm.

Once the graph is translated to a JT, the same VEA can be used to calculate the marginals.

For more information about translating graphs to JTs, please refer to [6, 10, 24, 26, 50].

PGMs have already a long history of use in database research [11, 19, 45, 47, 48, 51], albeit not for physical join algorithms.

## 3 THE GRAPHICAL JOIN (GJ)

GJ is a physical join algorithm; so all filters have already been applied and GJ only generates the join result on the filtered tables. For a given join query, GJ learns an exact MRF model from the data. The aim of GJ is to derive the factorized distribution from the normalized tables. Learning the factorized distribution is accomplished by scanning each table just once and calculating the exact frequencies. The de-normalization (join) in relational databases can be considered as the de-factorization in PGMs, but with the key difference that in PGMs what is de-factorized is the distributions, not data. Also, note that de-factorization is not a typical operation in PGMs. PGMs aim to provide efficient inference algorithms to answer queries over the factorized distribution. However, our take is that de-factorization can be beneficial when using PGMs in relational databases, where access to the flat full join result is needed. The questions are how can one de-factorize the MRFs efficiently (avoiding UIRs) to find the join result and at the same time how can one reduce the storage cost to store all needed information from which to produce the join result.

GJ's answer is to generate a summary of the join result without actually joining the tables with efficient algorithms. GJ can then store the summary in disk, and load and desummarize it when needed. The large space efficiency of GJ implies that even join queries with massive results can be handled, where all competing approaches fail. This scalability will be showcased in the experimental evaluation section.

This section shows how PGM fundamentals and algorithms (principled solutions for probabilistic inference) are used to generate

GFJS. By construction, GFJS is based on consecutive conditional frequencies. By building a PGM and running a VEA one time one can find all needed conditional frequencies needed for GFJS. Hence, with GJ, the physical join problem is mapped onto PGMs and tweaked algorithms are produced to deal with UIR (i.e., being a WOJA) and redundancy in the join result (i.e., being scalable, capable of dealing with massive joins, by generating GFJS).

GJ applies two modifications to the standard VEA: (i) It does not keep the frequency for all the combinations of variables' values in the potential functions. (ii) The output of the VEA in GJ is not the same marginal as in the standard VEA. GJ uses VEA to build a generative model which includes all the consecutive conditional potentials for generating GFJS.

## 3.1 An Overview of GJ

The whole process is shown in Figure 4. The user submits a join query. GJ builds a PGM (MRF) for the given query. The MRF has two components: qualitative and quantitative. The qualitative component is based on the query and the schema and includes only the graph $G(V, E)$ with no potentials. The quantitative component contains all the potentials. Each potential contains the exact frequencies per attribute distinct values. Calculating potentials simply requires scanning each table once, calculating these frequencies. Note that potentials may have been calculated for previous queries.

Note that JT creation is applied only on graphs with cycles and the aim of JT creation is to find a tree of maxcliques where the size of the largest maxclique in that tree is minimum in comparison to all other possible trees of maxcliques. Importantly, in our physical join problem, the number of nodes in the MRFs is usually small (just few nodes per involved attributes), and JT creation can be done manually as well, although the standard algorithms work well even with thousands of nodes. Once the JT with all potentials for the query is ready, GJ uses Algorithm 2 to build the generative model for GFJS based on an elimination order (recall trees have always at least a perfect elimination order).

During variable elimination, GJ prunes all MRF paths related to UIRs using a dynamic programming based approach with time complexity of the size of the largest potential function in the model. If the join graph is a tree then the largest potential comes from a single table, thus the complexity is in $O(M)$ where $M$ is the size of the largest potential and $N$ is the size of the biggest table. Note that $M \leq N$ always holds. If the join graph is a JT then the largest potential may come from several tables. In the worst-case scenario (which is very rare), let us assume we add all the nodes of the join graph in a single maxclique by adding all possible fill-in edges. In this case, we will show that the complexity is $O(M^{\rho})$.

Next, GJ uses Algorithm 3 to generate GFJS by using the generator output by Algorithm 2. The generation process is done in the reverse order of the elimination order. Here, the RLE generation over joins without doing the actual join is accomplished. Note that GJ does not go through the paths related to UIR tuples, and this is how it achieves WOJA status. Since GFJS only keeps frequencies, not raw data, the upper bound for GFJS's size is $O(M^{\rho})$.

After GFJS generation, it can be stored for later use. In the end, GFJS is desummarized whenever the full join result is needed. desummarization is straightforward, cf. Section 3.5.1.

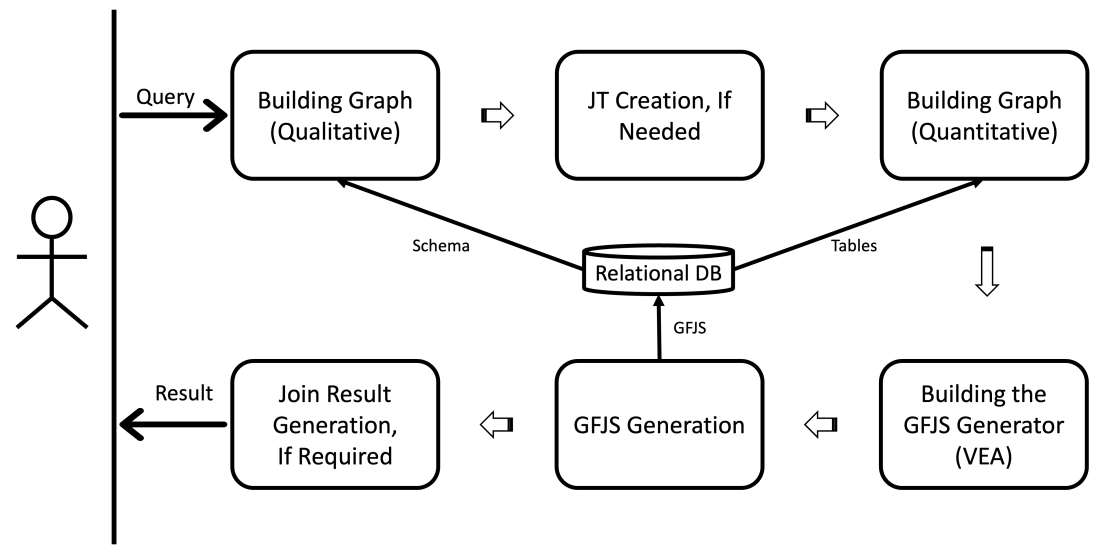

**Figure 4: Overview of GJ**

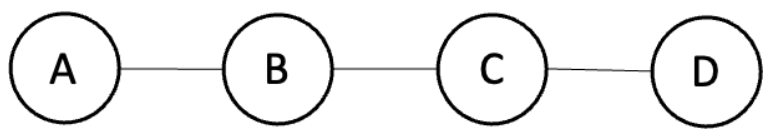

**Figure 5: Graph for the example join**

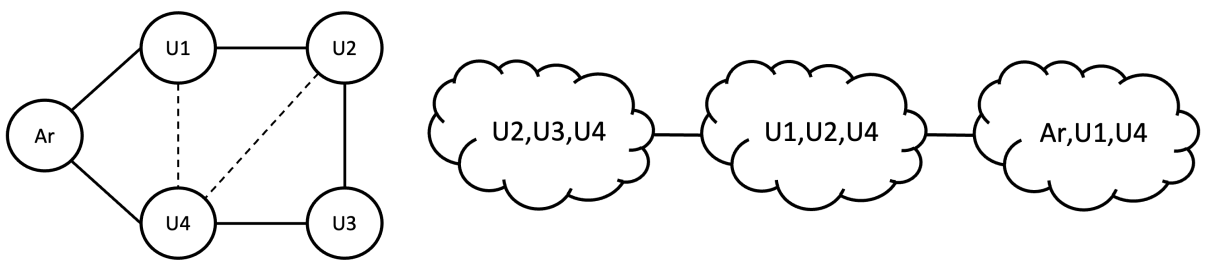

**Figure 6: a. Triangulated graph for lastFM_cyc b. Junction Tree with three maxcliques of size 3 for the graph**

Overall, GJ will be shown to have a complexity of $O(N^{\rho})$ to generate the join result in memory, and a complexity of $O(M^{\rho})$ when it is to store the result on disk.

## 3.2 Building MRFs for Join Queries

Although PGMs are considered 'heavy', our mapping of the join problem to PGMs results in a lightweight process. In general, learning a factorized distribution for a given joint distribution $p(x_V)$ is a heavy process. It requires finding all conditional independences among random variables and then use those to build an MRF. Furthermore, for each maxclique in the MRF, a potential function (factor) is learned and the product of all factors yields $p(x_V)$. However, in our case, all conditional independences among attributes are available because non-join attributes from different tables are conditionally independent from each other if the join attributes are observed. Moreover, learning the potential functions is straightforward because the data is already normalized and one can simply scan the tables to calculate the exact potentials (frequencies).

In qualitative learning, for each join attribute and non-join attribute (projection attribute) in the query, GJ adds a node in the graph. If two attributes are in the same table, an edge is added between their corresponding nodes in the graph. All involved attributes from the same table make a clique in the graph. Figure 5 shows the MRF for the natural join of the tables in Figure 1.

Quantitative learning starts after identifying the cliques and creating any junction trees. Cliques can be considered as hyper edges. In this step, GJ scans all the tables one time and calculates the potential functions (a.k.a factors) of all variables whose corresponding attributes come from the same table. The three potential functions of the graph in Figure 5 are shown in the top row of Figure 8. Each entry of a potential function is distinct and contains values for all the attributes in the same table and the frequency of the entry. Each table is scanned once and the complexity of learning the model fully is $O(N)$, where $N$ is the size of the largest table.

Note that a potential can be calculated on a table only once for all queries. In other words, quantitative learning can be offline which can reduce join-query processing time.

In practice, these potentials/factors are implemented with hash maps which give the frequencies of given keys in $O(1)$. The potentials are not necessarily un-conditional; they could be conditional in which case, nested hash maps are built for them. Needless to say, converting a joint distribution to a conditional one and vice versa needs to go through all the entries in the potential, so the complexity is equal to the size of the potential, $O(M)$.

#### 3.2.1 *Joining the Distributions (Potentials) for Cyclic Queries.*

Figure 6.a is for the lastFM_cyc query. Figure 6.b is for the JT derived from the cyclic graph. In this JT, there are maxcliques whose edges come from different tables, so we need to join (product) the potentials inside the maxcliques. For example the edge (*Ar, U1*) and (*Ar, U4*) come from different tables. To join the potentials and make the joint distribution (a single potential) for the maxclique, a novel WOJA algorithm is devised. This is shown in Algorithm 1. This algorithm is very similar to generic WOJAs from the literature (please refer to [17]), but the novelty is that it `joins potentials (distributions) rather than the data`.

Algorithm 1 is a recursive function getting a maxclique $C$ whose clique potentials come from different tables and generates a single potential that includes all the variables in $C$. In Algorithm 1, $V_C$ is the set of all variables in $C$, $O_C$ is the elimination order, $E_C$ contains all hyper edges (which represent the cliques coming from different tables) in $C$, $m$ is the number of potentials in $C$, $\Phi_C$ contains all potentials in $C$ and $i$ is the index of the variables to be considered. The algorithm starts with $i = 0$ and a recursion happens for each next variable. For each variable $v_i$ it finds the potentials that include $v_i$ as $\Phi'_C$ and those that do not include $v_i$ as $\Phi''_C$ in lines 4 and 5, respectively. Next, for each distinct value $k_i$, a shared value for $v_i$ among all potentials in $\Phi'_C$, the algorithm filters all the entries from all potentials in $\Phi'_C$ that do not have the value $k_i$ for their $v_i$ and makes new smaller potentials in $\Phi_{next}$ which contains only the entries with $k_i$ (in line 13). Then a new potential set $\Phi_C$ from the union of $\Phi_{next}$ and $\Phi''_C$ is made to be used to materialize the entries for the next variables (in line 14). If $v_i$ is the final variable, function **Bucket_Product** is called. This function calculates a product of all the entries in all the potentials in the last $\Phi_C$ and makes the final joint potential for $C$. Each entry has a frequency, so they are multiplied during the product calculation. For example, assume a triangle query having three potential functions coming from $T_1(A, B), T_2(B, C)$ and $T_3(C, A)$ which can be considered as a maxclique $(A, B, C)$. For the entries $(a_1, b_1, 5)$, $(b_1, c_1, 10)$ and $(c_1, a_1, 20)$ in the three potentials, an entry is added to the joint potential for the maxclique $(A, B, C)$ like $(a_1, b_1, c_1, 1000)$ which shows the frequency of $a_1, b_1, c_1$ in the join result of the three potentials.

**Asymptotic Complexity Analysis for Algorithm 1:** As explained, Algorithm 1 follows the same paradigm as all WOJAs, but the difference here is that Algorithm 1 joins the distributions (frequency tables or potentials). It is already been shown in all WOJA papers (such as [17, 34, 49]) that the WOJAs have the complexity of $O(N^\rho)$. Thus assuming that the largest potential in the maxclique is $M$, the complexity of Algorithm 1 is $O(M^\rho)$.

**Algorithm 1** A New WOJA algorithm to calculate the joint potential of all cliques in a given maxclique

For a given maxclique $C$:
$V_C = \{v_1, v_2, ..., v_n\}$ is the set of variables in $C$.
$O_C$ is a set of all $v \in V_C$ but with the elimination order.
$E_C = \{e_1, e_2, ..., e_m\}$ is the set of hyper edges which makes the cliques in $C$. Each $e$ contains a subset of $V_C$
$\Phi_C = \{\phi_1, \phi_2, ..., \phi_m\}$ is the set of to-be-joined potentials in $C$ per hyper edge.
$i$ is the index for the current variable.

1: **procedure** Potential-Join($i, \Phi_C$)
2: **if** $i < n$ **then**
3: $v_i \leftarrow O_C[i]$ ▷ $v_i$ is the variable to be eliminated
4: $\Phi'_C \leftarrow \{\phi_j \in \Phi_C | v_i \in e_j\}$ ▷ All the potentials that include $v_i$
5: $\Phi''_C \leftarrow \Phi_C - \Phi'_C$ ▷ All the potentials that exclude $v_i$
6: **for each** $k_i \in \bigcap_{\phi_j \in \Phi'_C} \pi_{v_i}(\phi_j)$ **do** ▷ $k_i$ is the shared value
7: $\Phi_{next} \leftarrow \{\sigma_{v_i=k_i}(\phi_j) | \phi_j \in \Phi'_C\}$ ▷ $\sigma$, selection operator
8: **Potential-Join**$(i+1, \Phi_{next} \bigcup \Phi''_C)$ ▷ The recursive part
9: **end for**
10: **else**
11: **Bucket_Product** $(\bigtimes_{\phi_j \in \Phi_C} \phi_j)$ ▷ Potential product with freqs
12: **end if**
13: **end procedure**

## 3.3 GFJS Generator: Inference on Trees

An algorithm for creating a GFJS generator on trees is now provided.

Algorithm 2 calculates all conditional and un-conditional potentials (as a GFJS generator) in an efficient way based on the dynamic programming of VEA. Note, that all potentials in the query MRF and the new potentials generated by Algorithm 2 exclude the entries with frequencies smaller than 1. This helps GJ to avoid UIRs when generating GFJS. In Algorithm 2, $V, O, E$ and $\Phi$ are the same as in Algorithm 1, but for the full query graph (not for a single maxclique). $P$ contains the parents for $v_i \in V$. $\Psi$ is the GFJS generator which contains a marginal for the root, and conditional frequency tables for other variables based on the elimination order in $O$. Since trees always have a perfect elimination order, which starts from the leaves to the root, $O$ can be obtained by eliminating the variables from the leaves to the root. The loop in lines 2-11 of Algorithm 2 eliminates the variables one by one, using sum-product operations. Lines 4 and 5 find the potentials that include the current to-be-eliminated variable $v_i$ as $\Phi'$ and those that exclude $v_i$ as $\Phi''$. In line 6, all the potentials in $\Phi'$ are multiplied and stored in $\phi_\alpha$ (the product part of the sum-product operation). Lines 7 conditionalize $\phi_\alpha$ on $v_i$'s parent $P_{v_i}$ as $\psi_{v_i|P_{v_i}}$ and line 15 adds it in $\Psi$, which is the GFJS generator. Line 16 does the summing out part of the sum-product operation. In this step, $v_i$ is summed out from the product of all potentials that include $v_i$ and is shown as $\phi_\beta$. $\phi_\beta$ is the potential for the remaining variables in $\phi_\alpha$ after eliminating $v_i$. Line 17 updates $\Phi$ by removing the potentials that exist in $\Phi'$ and adding the new potential $\phi_\beta$. Once all variables are eliminated, all potentials for the root are also multiplied to have a single marginal for the root. This marginal for the root will be the starting point when generating GFJS. It is worth mentioning that the sum of all the frequencies in the root marginal is exactly equal to the join size.

To make Algorithm 2 clearer, an example of building the GFJS generator is provided in the following.

**Algorithm 2** Building the generative model to generate the GFJS

For a given query graph $G$:
$V = \{v_1, v_2, ..., v_n\}$ is the set of variables in $G$.
$O$ is a set of all $v \in V$ but with the elimination order.
$P$ includes the parents of each variable in $V$.
$E = \{e_1, e_2, ..., e_m\}$ is the set of hyper edges which makes the cliques in $G$. Each $e$ contains a subset of $V$.
$\Phi = \{\phi_1, \phi_2, ..., \phi_m\}$ is the set of potentials in $G$ per hyper edge.
$\Psi$ is the GFJS generator.

**procedure** BUILD_GFJS_GENERATOR
  **for** $i = 0$ to $n$ **do** ▷ The loop is run per variable
    $v_i \leftarrow O[i]$ ▷ The current variable to eliminate
    $\Phi' \leftarrow \{\phi_j \in \Phi | v_i \in e_j\}$ ▷ All the potentials which includes $v_i$
    $\Phi'' \leftarrow \Phi - \Phi'$ ▷ All the potentials which excludes $v_i$
    $\phi_\alpha \leftarrow \prod_{\phi \in \Phi'} \phi$ ▷ Product of all the related potentials
    $\psi_{v_i|P_{v_i}} \leftarrow conditionalize(\phi_\alpha, P_{v_i})$ ▷ condition on $v_i$'s parent
    $\Psi \leftarrow \Psi \bigcup \psi_{v_i|P_{v_i}}$
    $\phi_\beta \leftarrow \sum_{v_i} \phi_\alpha$ ▷ Summing out $v_i$ from the product
    $\Phi \leftarrow \Phi'' \bigcup \{\phi_\beta\}$ ▷ The new $\Phi$ includes potentials with no $v_i$
  **end for**
  $\Psi \leftarrow \Psi \bigcup \prod_{\phi \in \Phi} \phi$ ▷ Unconditional potential for the root
**end procedure**

**Example.** Consider the graph in Figure 5 (our running example). The potentials for the edges are shown in the top part row of Figure 8. These potentials are calculated by scanning the tables in Figure 1. Based on Equation 2, the joint distribution of the three tables is:

$$p(A, B, C, D) \propto \phi(A, B)\phi(B, C)\phi(C, D) \quad (5)$$

Given an elimination order $O = \{D, C, B, A\}$, Algorithm 2 will eliminate the variables $D, C$ and $B$ respectively and will find the marginal for $A$ which is the root according to $O$.

$$p(A) \propto \sum_{D,C,B} \phi(A, B)\phi(B, C)\phi(C, D) \quad (6)$$

Based on the distributive law, this can be represented as:

$$p(A) \propto \sum_{C,B} \phi(A, B)\phi(B, C) \sum_{D} \phi(C, D) \quad (7)$$

Now we have the full tree and an empty GFJS generator, please refer to the first row of Figure 7. The black graph is the main graph and the orange graph represents the GFJS generator.

After summing out variable $D$, a new potential (factor) is added for $C$. This new factor is shown in the middle row of Figure 8 (the right most factor).

$$p(A) \propto \sum_{B} \phi(A, B) \sum_{C} \phi(B, C)\phi_{\not{D},C}(C) \quad (8)$$

While eliminating $D$, a conditional $\phi(D|C)$ is also calculated to be added in our GFJS generator. This conditional potential gives the frequency of $D$ values given $C$ values over the sub-tree starting from $C$ to the leaves ($D$). Since $D$ is a leaf, the sub-tree contains only the edge from $C$ to $D$. The conditional potential is shown in the bottom row of Figure 8 (the right most one). Each entry of the conditional factors contains the variable(s) $v_j$ on which the condition is on ($C$ in our example), the dependent variable/variables $v_i$ ($D$ in our example), a *bucket* value which shows the local frequency of $v_i$ given $v_j$ in the table containing both $v_j$ and $v_i$ (table 3 in our example), and a $fac$ value which shows the frequency of $v_i$ in the factor coming from the children of $v_i$ (there are no children in this step for $D$). For the conditional factors where the dependent variable is a leaf, $fac$ values are 1s. The conditional factors are implemented as nested hash maps. The graph for this conditional factor is shown in the second row of Figure 7.

Next, $C$ is eliminated. After eliminating $C$, we have

$$p(A) \propto \sum_{B} \phi(A, B)\phi_{\not{D},\not{C},B}(B) \quad (9)$$

This operation will add a factor for $B$ (refer to the middle factor in the middle part of Figure 8), and a conditional factor $\phi(C|B)$ in the GFJS generator as shown in the bottom part of Figure 8. The new conditional factor is shown in the third row of Figure 7, and $C$ is eliminated from the main graph. In fact, $\phi(C|B)$ gives the frequency of $C$ given $B$ over the sub-tree from $B$ to the leaves ($D$).

Next, $B$ is eliminated. This will add a factor for $A$, as shown in the middle part of Figure 8 (the left most one), and a conditional factor $\phi(B|A)$ in the GFJS as shown in the bottom part of Figure 8 (the left most one). $\phi(B|A)$ gives the frequency of $B$ given $A$ over the whole tree. The changes in the main graph and the GFJS generator are shown in the lowest row of Figure 7.

$$p(A) \propto \phi_{\not{D},\not{C},\not{B},A}(A) \quad (10)$$

The final (un-conditional) factor is added to the GFJS generator at the end. The sum of all the frequencies in $p(A)$ give the join size.

As mentioned, WOJAs need a variable order when generating the join tuples, which they choose randomly. Mapping the join problem to PGMs, allows one to actually show that choosing a bad order can significantly affect performance. For instance, assume $B$ is eliminated first from Equation 6. In this case, first the product of $\phi(A, B)$ and $\phi(B, C)$ as $\phi(A, B, C)$ should be calculated and then $B$ is summed out. In other words, a new edge between $A$ and $C$ must be added to preserve the dependency between $A$ and $C$ in the absence of $B$. Since eliminating $B$ involves two potentials and three variables, it is less efficient than eliminating $D$ from Equation 6 with one potential and two variables.

*3.3.1 Asymptotic Complexity of Inference on Trees.* Since trees have a perfect elimination order and each variable has a single parent, each elimination involves a single potential with two variables. This potential comes from a single table and there is no repetition in the entries of the potentials. Thus, given that the size of the biggest potential is $M$, the complexity of eliminating all variables is $O(M)$.

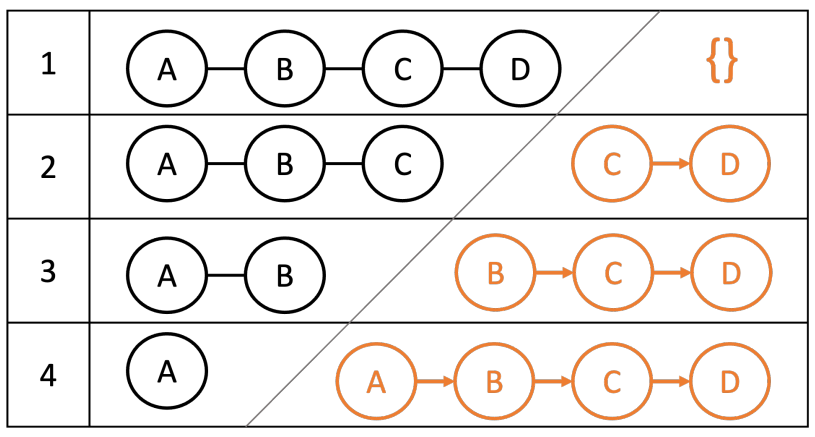

Figure 7: Join graph and the GFJS generator

| Potential for table 1 | | |
|---|---|---|
| A | B | freq |
| a0 | b0 | 3 |
| a1 | b1 | 2 |
| a2 | b1 | 1 |
| a3 | b3 | 2 |
| | b4 | 4 |

| Potential for table 2 | | |
|---|---|---|
| B | C | freq |
| b0 | c0 | 2 |
| b1 | c0 | 3 |
| b2 | c1 | 3 |
| b3 | c2 | 1 |
| b4 | c3 | 2 |
| | c4 | 1 |

| Potential for table 3 | | |
|---|---|---|
| C | D | freq |
| c1 | d0 | 4 |
| c2 | d2 | 4 |
| c3 | d3 | 2 |
| c4 | d4 | 2 |

| $\phi_{B\rightarrow A}(A)$ | |
|---|---|
| A | freq |
| a3 | 32 |

| $\phi_{C\rightarrow B}(B)$ | |
|---|---|
| B | freq |
| b2 | 12 |
| b3 | 4 |
| b4 | 6 |

| $\phi_{D\rightarrow C}(C)$ | |
|---|---|
| C | freq |
| c1 | 4 |
| c2 | 4 |
| c3 | 2 |
| c4 | 2 |

| $\phi(B\|A)$ | | | |
|---|---|---|---|
| A | B | bucket | fac |
| a3 | b3 | 2 | 4 |
| | b4 | 4 | 6 |

| $\phi(C\|B)$ | | | |
|---|---|---|---|
| B | C | bucket | fac |
| b2 | c1 | 3 | 4 |
| b3 | c2 | 1 | 4 |
| b4 | c3 | 2 | 2 |
| | c4 | 1 | 2 |

| $\phi(D\|C)$ | | | |
|---|---|---|---|
| C | D | bucket | fac |
| c1 | d0 | 4 | 1 |
| c2 | d2 | 4 | 1 |
| c3 | d3 | 2 | 1 |
| c4 | d4 | 2 | 1 |

Figure 8: Potentials, factors/messages and conditional factors for the graph in Figure 5

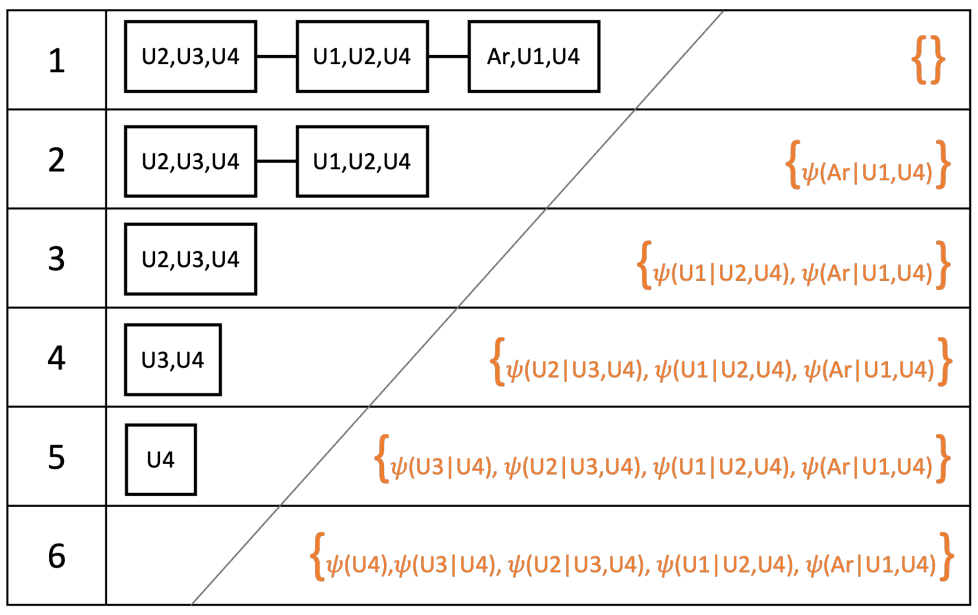

Figure 9: Inference on the JT of Figure 6

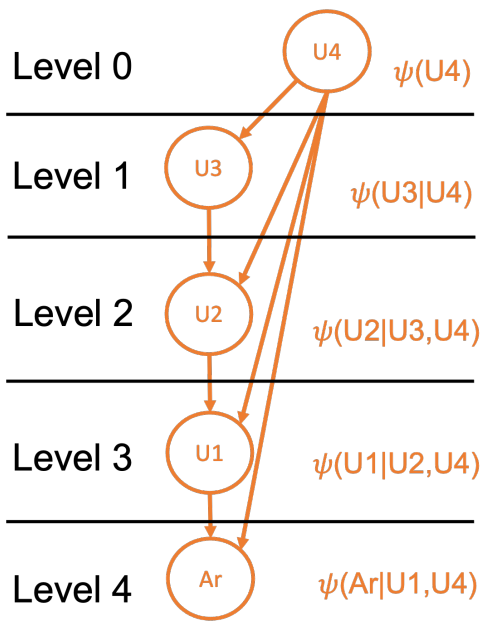

Figure 10: The DAG for the GFJS generator in Figure 9

## 3.4 GFJS Generator: Inference on Graphs

If the query graph is not a tree, it must be translated to a new structure having the same properties (e.g., R.I.P) as trees. Hence, a JT (a tree of maxcliques) is created using standard algorithms as explained in the Background section. Once the JT is created, the above inference algorithms run over JTs.

Concretely, query lastFM_cyc from our experiments is cyclic yielding a cyclic graph. One of the possible JTs for that query is shown in Figure 6. The joint distribution for the JT in Figure 6 is:

$$p(Ar, U1, U2, U3, U4) \propto \phi(U2, U3, U4)\phi(U1, U2, U4)\phi(Ar, U1, U4) \tag{11}$$

Given an elimination order like $O = \{Ar, U1, U2, U3, U4\}$, the variables are summed out from Equation 11 one by one. Below all the steps of eliminating the variables are shown. For each elimination step, the changes applied to the main JT and the GFJS generator are shown in Figure 9. The difference here is that the conditions are on composite keys (the separator sets).

$$p(U4) \propto \sum_{U1,U2,U3} \phi(U2, U3, U4)\phi(U1, U2, U4) \sum_{Ar} \phi(Ar, U1, U4)$$

$$p(U4) \propto \sum_{U2,U3} \phi(U2, U3, U4) \sum_{U1} \phi(U1, U2, U4)\phi_{\cancel{Ar},U1,U4}(U1, U4)$$

$$p(U4) \propto \sum_{U2,U3} \phi(U2, U3, U4)\phi_{\cancel{Ar,U1},U2,U4}(U2, U4)$$

$$p(U4) \propto \sum_{U3} \phi_{\cancel{Ar,U1,U2},U3,U4}(U3, U4)$$

$$p(U4) \propto \phi_{\cancel{Ar,U1,U2,U3},U4}(U4)$$

### 3.4.1 *Asymptotic Complexity of Inference on Graphs.*

The asymptotic complexity of Algorithm 2 on graphs is different from that on trees. In a JT, it is possible to have a maxclique with some potentials (cliques) that come from different tables. In this case, we use the new Algorithm 1 to join potentials (frequency tables, not data tables) and make a single potential for the maxclique. Recall that the complexity of the inference on trees was equal to the size of the largest potential ($O(M)$), but here the largest potential can be larger as the maxclique may involve several potentials. In the worst-case (which is very rare), all variables are added into a single maxclique. In this case, joining all potentials of the graph with Algorithm 1 is $O(M^{\rho})$. This is because Algorithm 1 is a WOJA.

## 3.5 Summary Generation

The GFJS generator $\Psi$ for chains is a chain, for trees is a tree and for JTs it is a directed acyclic graph (DAG).

The orange graph in the lowest row of Figure 7 is the GFJS generator for our three-table running example. This generator has 4 levels and the root is $A$. The GFJS generator for our cyclic query is shown in Figure 10 with 5 levels and the root is $U4$.

The summary for the root is already ready, $\psi_0$. $\psi_0$ is the last (root) potential from the inference and keeps the frequency of the root variable over the join result.

GFJS for other variables is generated recursively, per level of the generator. In Algorithm 3, for each entry in the $\psi_0$ of the generator, the recursive function of Algorithm 4 is called which generates the summaries per level of the generator.

Note, in our examples, we have a variable per level, but it is possible to have several variables in a level. For example, if $O$ is $\{D, C, A, B\}$ to do inference in our three-table example, $B$ would be the root and it would have two children $A$ and $C$.

In line 2 of Algorithm 4, all potentials in level $i$ of the generator are conditioned on the already observed values in their parents, and are added to $\Psi'$. $keys$ is a map from the variable names to their values that are already observed in upper levels. The operator [] makes the conditional $\psi$s, unconditional by applying the conditions using the observed values in $keys$. $\Psi'$ holds the unconditional potentials for all the variables in the same level. If there are more than one variable in the current level, Algorithm 4 calculates the Cartesian product of values. In the Cartesian product, the values for all variables are combined, the $bucket$ values are multiplied together and $freq$ values are also multiplied. The result is stored in $\mathcal{R}$. For each row $r$ in $\mathcal{R}$, the recursive function is called again to add the summaries for the next variables. In line 5, the already observed values are combined with the values in $r$ for the current level variables. Also, the $bucket$ value from the upper level and the bucket value for $r$ are multiplied. Line 7 adds the summary for the variables in the current level $i$ into $s_i$ by multiplying the new bucket value with the $freq$ values for each $r$. If it has not reached the last level, the recursive function is called again in line 9. GFJS for our 3-table running example is shown in Figure 2. The first column ($A$) of GFJS comes from $\psi_0$, the second column ($B$) is generated based on the observed values for the root $A$, then column for $C$ is generated based on the observed values for $B$, and at the end the column for $D$ is generated based on the observed values for $C$.

Note, since $\psi_0$ does not include any unneeded value for the root, and since we do not have any entry in $\psi \in \Psi$ with zero frequencies, GFJS generation will avoid UIR; hence, GJ is a WOJA.

To handle projection operations, if a node in the graph is not the projection list (i.e., an output attribute), one can ignore the summary for that node and simply pass to the next level. This happens when some of join attributes are not in the select clause. However, Section 3.7 will explain how one can efficiently apply projections on graphs before starting the summary generation, by deleting those variables from the models. Consequently, the attributes that are not in the output will be deleted from the GFJS generator.

*3.5.1 Asymptotic Complexity Analysis for GFJS Generation .* The asymptotic complexity for GFJS generation is $O(M^\rho)$ since it does not generate any information that is not in the join result and it adds an entry in the GFJS per distinct value.

## 3.6 Join Result Generation (Desummarization)

One of the advantages of RLE ( in our case, the summary columns of GFJS) is that RLE can be stored/retrieved/desummarized straightforwardly. For each column of the GFJS, GJ creates a CSV file.

**Algorithm 3** Generates the GFJS given a generator $\Psi$

$\psi_0$ is the root potential for the GFJS generator $\Psi$.
$S = \{s_0, s_1, ..., s_{m-1}\}$ is the set of summaries per GFJS column.
$m$ is the number of levels in $\Psi$.
$root$ is the variable name for the root.

**procedure** $gen_GFJS()$
  **for each** $entry$ in $\psi_0$ **do** ▷ Each entry is $(value, freq)$
    $keys \leftarrow \{root, entry.value\}$ ▷ root value, observed
    $s_0 \leftarrow s_0 \cup (entry.value, entry.freq)$ ▷ GFJS for the root
    $rec_GFJS(1, 1, keys)$ ▷ To find values for other variables
  **end for**
  $S \leftarrow \{s_0, s_1, ..., s_{m-1}\}$ ▷ $S$ is the final GFJS
  store $S$ on disk, if needed
**end procedure**

**Algorithm 4** Recursive algorithm used in Algorithm 3

$\Psi$ is the GFJS generator.
$\Psi_i$ gives all the $\psi$s in level $i$.
$s_1, s_2, ..., s_{m-1}$ are the summaries per level.
$m$ is the number of levels in $\Phi$.
$keys$ is a map of the variable names to their chosen values.
$i$ is the current level index.
$p_bucket$ is the bucket size from the upper level.

1: **procedure** $rec_GFJS(i, p_bucket, keys)$
2:   $\Psi' \leftarrow \bigcup_{\psi \in \Psi_i} \psi[keys]$ ▷ $keys$ contains the observed values
3:   $\mathcal{R} \leftarrow \bigtimes_{\psi \in \Psi'} \psi$ ▷ $\psi \in \Psi'$ is unconditional
4:   **for each** $row$ in $\mathcal{R}$ **do**
5:     $keys_{new} \leftarrow keys \bigcup row.keys$ ▷ New values for next $keys$
6:     $bucket_{new} \leftarrow p_bucket \times row.bucket$
7:     $s_i \leftarrow s_i \cup (row.keys, bucket_{new} \times row.freq)$
8:     **if** $i < m - 1$ **then** ▷ If any variable remains then recur
9:       $rec_GFJS(i + 1, bucket_{new}, keys_{new})$
10:     **end if**
11:   **end for**
12: **end procedure**

Desummarization starts from the top of the column (let us say $s_1$) in $S$ and replaces any $(v, freq)$ pair with $freq$ of $v$ values, where $v$ is the value and $freq$ is the value's frequency. The cost of desummarization is exactly the same as the join size $|Q|$. Desummarization is avoided if the join result is to be stored on disk.

## 3.7 Early Projections

Not all join attributes are output attributes. In other words, we could have nodes in the graph that are not supposed to be generated in the join result. The idea of early projection is that, before starting the GFJS generation, one can delete these nodes from the join graph and the GFJS generator. When finding the elimination order, one could find two elimination orders $O$ and $O'$, where $O$ is related to the output (projection) attributes and $O'$ is for the join attributes that are not output attributes. $O'$ is always considered before $O$. First, the unneeded variables based on $O'$ are deleted. Then the others are eliminated, based on $O$. When "a variable is eliminated", it means that a factor for its parent is calculated (recall the second row of Figure 8) and also a conditional factor is calculated (recall the third row of Figure 8). But when "a node is deleted", it means that the conditional factors for the node are not generated and the node is deleted from the graph; but, the factor for its parent is still calculated. Projections do not affect the asymptotic complexity of the inference which is still $O(M^{\rho})$, in the worst case. GJ is, to our knowledge, the first WOJA able to apply early projections.

## 3.8 Overall Asymptotic Complexity of GJ

Overall, the four main operations of GJ are: i) scanning the tables (when potentials are not already available). This is accomplished in $O(N)$ for tree queries and $O(M^{\rho})$ for JTs; ii) running inference on the query graph to build the GFJS generator. This operation on trees is done in $O(M)$. On graphs, it is equal to the size of the largest potential for maxcliques. In the worst (rare) case, when creating JTs, we add all nodes in a single maxclique. Then the asymptotic complexity is $O(M^{\rho})$; iii) generating GFJS, which needs to traverse the graph one time. GFJS generation is carried out in $O(M^{\rho})$. Note that summary generation is performed on graphs whose unneeded intermediate paths have already been pruned. Moreover, since repeated values are replaced with frequencies, the summary is typically much smaller than the join result; iv) desummarizing, whose cost is equal to the join size $|Q|$ ($|Q| \leq N^{\rho}$).

Note, if we wish to store the join result on disk, GJ does not pay the cost for desummarization. In the worst-case, GJ is asymptotically as fast as any WOJA. But, its storage cost is significantly smaller, since GJ stores only the join summary.

# 4 EXPERIMENTAL EVALUATION

**Setup and Baselines**. We study the performance of GJ against PostgreSQL (PSQL), MonetDB, and the WOJA in Umbra , hereafter referred to as "Umbra" which is a state-of-the-art WOJA. We set Umbra to use its WOJA algorithm ($Umbra_{EAG}$) as the aim is to use it to compare GJ against WOJAs. These competitors are full-fledged systems, whereas GJ is a C++ code library working with CSVs (one per table in the join and possibly one storing the join result). Systems incur additional overheads when executing a join. Thus, to ensure a fair comparison, we obtained (from the authors of Umbra) a version that works with CSV files (e.g., escaping other system overheads, such as dealing with ACID properties, logging, etc). [1]. Thus, two versions of Umbra are used in the experiments, one as an RDBMS and another one which works with CSV files (like GJ). Also, please note that the aforementioned overheads in full-fledged RDBMSs represent typically a small percentage of execution time, whereas the differences between GJ and them are large, as will be shown later.

We evaluate under two use scenarios. The first (compute-and-reuse) scenario, needs accounting the costs for executing joins, storing results on disk and later reloading results into memory. In this scenario, GJ stores/reloads GFJS on/from disk, and then desummarizes it as a flat join result (whereas PSQL, MonetDB and Umbra store/reload the flat join result). We also compare the storage cost of GJ and other competitors.

In the second "compute-and-forget" scenario, GJ generates GFJS and from it then generates the flat join result. Furthermore, we show the PGM building time for GJ and also show the benefits of GJ when pre-building the PGM for frequent joins.

We use a server with one Intel Core™ i5-8500 3.00GHz CPU, with 32GB RAM, and 1TB SSD disk. We run each query 5 times and then report averages. Parallelism on MonetDB, PSQL, Umbra and GJ is disabled in an effort to compare solutions without spending more resources to address inefficiencies. We have also developed a parallel version of GJ. Parallel GJ is intricate and its presentation requires delving into parallel PGM algorithms and our extensions. Hence, for readability and space purposes, we omit here relevant discussions and evaluation. Suffice it to say, with parallelism turned on, GJ exhibits similar performance gains as reported here. Nonetheless, readers can find relevant discussions and evaluation results with parallelism turned on versus the above baselines here [2].

**Data:** We use the JOB [27] and TPCH benchmarks[3] (with scaling factor 1), as well as lastFM[4], a real-world data set [8]. MonetDB builds its own indices. For PSQL, we pre-built B-trees on all join attributes before running queries. Umbra builds its own tries.

**Queries:** We use 4 JOB queries, 5 lastFM queries, and 2 TPCH queries. Their description is omitted for space reasons - please find their details at https://github.com/shanghoosh1/Graphical_join/blob/main/queries.txt. When selecting our queries for experiments we wanted to bring out salient features of joins and how they impact GJ and the competitors. JOB queries include some many-to-many relations. Hence they have more redundancy in the join result and UIRs. lastFM queries have less redundancy in the result, but suffer from higher UIRs. TPCH queries occupy the other extreme, being foreign key (FK) joins, having low redundancy in the join result and there are no UIRs. Since we focus on physical joins, all non-join predicates (filters), aggregations and group-by statements are removed. The join sizes per query are listed in Table 1 where sizes range from ca. 5 million tuples to ca. 146 billion tuples.

## 4.1 Results for Query Time and Space

**The compute-and-reuse scenario**. Here, we show the key costs for: i) running the join and storing the result on disk, ii) loading time of the result into memory, and iii) storage. Time costs to run and store the join result on disk are shown in Table 2. Results with '>' mean that after a certain time the database crashed (typically, running out of storage - exceeding the available 1TB). The "-" means that after exceeding 1TB, Umbra-CSV neither crashed nor finished the job. The results show that GJ is always better than competitors, except for the FK TPCH joins (as expected). Since GJ does not store the actual join results (but GFJS), the speedup is drastic: up to 820X faster than PSQL; more than 717X faster than MonetDB; up to 165X faster than Umbra; and 94X faster than Umbra-CSV. Table 2 shows that the idea of calculating the summary of the join result without computing the join result, can have a significant impact on join performance. MonetDB is always better than PSQL when the join results are supposed to be stored on disk, likely as a result of the columnar nature of MonetDB. FK queries have less redundancy, no UIRs and also their join sizes are small; hence, GJ is worse than MonetDB and Umbra. None of the competitors could run JOB_D due to its large join size, exceeding 1TB with PSQL/MonetDB/Umbra/Umbra-CSV. GJ managed JOB_D in 51.8 seconds with a meagre storage cost of 260MB! Queries like JOB_D showcase the scalability afforded by GJ.

[1] We wish to sincerely thank the Umbra team (especially M. Freitag) for providing executables for Umbra and helping us run Umbra with various data and queries.

[2] https://github.com/shanghoosh1/Parallel_Graphical_Join.
[3] http://tpc.org/
[4] https://grouplens.org/datasets/hetrec-2011/

**Table 1: Join sizes per query**

| JOB_A | JOB_B | JOB_C | JOB_D | lastFM_A1 | lastFM_B | lastFM_cyc | FK_A | FK_B |
|---|---|---|---|---|---|---|---|---|
| 444,340,632 | 31,588,599,792 | 1,037,051,092 | 146,527,949,388 | 61,664,382 | 12,538,289,270 | 706,351,348 | 6,001,194 | 5,760,999 |

**Table 2: Time cost in seconds for generating and storing the join result in disk (GJ stores the GFJS)**

| | JOB_A | JOB_B | JOB_C | JOB_D | lastFM_A1 | lastFM_B | lastFM_cyclic | FK_A | FK_B |
|---|---|---|---|---|---|---|---|---|---|
| GJ | **8.8** | **27.6** | **45.2** | **51.8** | **5.8** | **80.8** | **9.8** | 2.49 | 5.12 |
| Umbra | 36.2 | 4,560 | 140.8 | >2,664 | 7.1 | 1,792 | 97.6 | 1.68 | **2.71** |
| Umbra-CSV | 23.7 | 2,610 | 157.6 | - | 11.9 | 423 | 154.7 | 1.82 | 2.74 |
| MonetDB | 42.3 | >19,794 | 480.9 | >4,537 | 8.4 | 6,810 | 1,639 | **0.91** | 2.85 |
| PSQL | 324.5 | 22,633 | 808 | >26,533 | 61 | 11,508 | 862 | 7.3 | 8.67 |

**Table 3: Time cost in seconds for loading the results into memory (GJ loads the summary and desummarizes it in the memory)**

| | JOB_A | JOB_B | JOB_C | JOB_D | lastFM_A1 | lastFM_B | lastFM_cyclic | FK_A | FK_B |
|---|---|---|---|---|---|---|---|---|---|
| GJ | **0.17** | **26.7** | **3.7** | **264** | 3.4 | **83.4** | **1.9** | **0.004** | **0.007** |
| Umbra | 0.58 | 3,156 | 139.8 | - | **0.72** | 2,016 | 118.8 | 0.056 | 0.055 |
| Umbra-CSV | 15.96 | 2,731 | 113.2 | - | 10.2 | 2,418 | 136.7 | 0.53 | 0.57 |
| MonetDB | 1.34 | - | 9.4 | - | 0.99 | 763 | 10.1 | 0.046 | 0.044 |
| PSQL | 60.6 | 7,063 | 232 | - | 6.4 | 3,458 | 203 | 0.91 | 0.39 |

**Table 4: Storage cost in MBs**

| | JOB_A | JOB_B | JOB_C | JOB_D | lastFM_A1 | lastFM_B | lastFM_cyclic | FK_A | FK_B |
|---|---|---|---|---|---|---|---|---|---|
| GJ | **4.4** | **50.8** | **351** | **260** | **312** | **5,529** | **89** | **0.0024** | **0.4** |
| Umbra | 7,639 | 994,099 | 49,624 | - | 3,993 | 777,206 | 11,195 | 189 | 181 |
| Umbra-CSV | 3,174 | 461,824 | 23,040 | - | 2,867 | 592,998 | 32,870 | 131.2 | 120.3 |
| MonetDB | 1,976 | - | 11,980 | - | 1,845 | 384,102 | 21,640 | 92 | 148 |
| PSQL | 15,360 | 1,091,584 | 44,032 | - | 3,068 | 623,616 | 35,148 | 247 | 236 |

Time costs for loading the join result from disk are shown in Table 3. GJ needs to load GFJS into memory and desummarize it. Results in Table 3 show that GJ is always faster that PSQL in preparing the join result in memory up to 356X (with query JOB_A). GJ is also always better than MonetDB (up to 11.5X with query FK_A) except for lastFM_A1. GJ is always better than Umbra (up to 118X with query JOB_B) except for lastFM_A1 and lastFM_cyc. GJ is always faster than Umbra-CSV (up to 132X with query FK_A). Furthermore, as CPUs are getting faster, much faster than disks, loading a small GFJS and desummarizing it will be increasingly beneficial than loading full join results from disk.

The reason that Umbra-CSV is slower than Umbra in loading results into memory is that Umbra-CSV needs to parse strings from CSV files, which takes significant time. Note that GJ also parses the strings in CSV files. This overhead is not incurred in an RDBMS which may make GJ even faster if implemented within an RDBMS.

In general, GJ shines when there are UIRs, high redundancy in the results, and join sizes are big.

Storage costs per query are shown in Table 4. For all queries GJ is dramatically more efficient, up to 21,488X (with query JOB_B) better than PSQL, up to 38,333X (with query FK_A) better than MonetDB, and up to 78,750X (with query FK_A) better than Umbra and up to 54,666X (with query FK_A) better than Umbra-CSV. MonetDB could not store the result for JOB_B and JOB_D, and all competitors could not finish JOB_D.

**The compute-and-forget scenario**. The times to compute the join result in memory are shown in Table 5. GJ is always better than PSQL and MonetDB, especially on lastFM queries, as expected. The impact of GJ for FK joins is less, again as expected. GJ is better than Umbra for all many-to-many queries except for JOB_A and JOB_C. GJ is faster than PSQL by up to 64X (with query lastFM_B); faster than MonetDB by more than 388X (with query JOB_B); and faster than Umbra by more than 6X (with JOB_B and lastFM_B). The results for Umbra-CSV are almost the same as Umbra as we do not generate, store and reload the CSV files for in-memory runs. Generally speaking, wherever the join size is larger, the efficiency gain from GJ's summarization/desummarization is higher.

Table 6 shows the percentage of GJ's in-memory runtime which is spent on building the PGM - actually, this refers to the cost to compute the potentials (frequency tables), as creating the graph structure is trivial. For example, 99% of the time for query FK_A, 62% for query JOB_A, and 0.52% for FK_B are for PGM building. We refer to these queries in particular as GJ performs worse for these queries, cf. Table 5. Noting this, one could pre-build the frequency tables and keep them in memory for frequently used tables for joins. The pre-built frequency tables can also be used for other purposes (e.g. for cardinality estimation). Prebuilding PGMs for frequent joins can actually make GJ competitive, even for FK-joins.

**Table 5: Time cost in seconds for running the joins in memory**

| | JOB_A | JOB_B | JOB_C | JOB_D | lastFM_A1 | lastFM_B | lastFM_cyclic | FK_A | FK_B |
|---|---|---|---|---|---|---|---|---|---|
| GJ | 8.9 | **51** | 43.9 | **302** | **0.62** | **33.14** | **9.96** | 2.51 | 5.12 |
| Umbra | **4.7** | 305 | **39.2** | 1620 | 1.02 | 221.2 | 29.61 | 1.17 | **2.15** |
| MonetDB | 38.5 | >19,776 | 411 | >4,155 | 3.13 | 3,466 | 1,713 | **0.66** | 2.69 |
| PSQL | 40.6 | 2,219 | 109 | 10,660 | 7.51 | 2,101 | 253 | 4.1 | 5.28 |

**Table 6: The percentage of in-memory running times spent on building PGMs by GJ**

| JOB_A | JOB_B | JOB_C | JOB_D | lastFM_A1 | lastFM_B | lastFM_cyclic | FK_A | FK_B |
|---|---|---|---|---|---|---|---|---|
| 62% | 31% | 49% | 9% | 9% | 0.3% | 1% | 99% | 52% |

## 4.2 Sensitivity Analysis: UIR and Redundancy

Query lastFM_A1 was included since it has UIR and result redundancy. The query is about users and their friendship and it joins tables *user_artists*, *user_friends* and *user_artists*. We now evaluate a variation, lastFM_A2, to see how larger UIR can affect performance. lastFM_A2 entails users and their relation with their friends of friends. lastFM_A2 joins tables *user_artists*, *user_friends*, *user_friends* and *user_artists*. The additional join operation increases UIR. Furthermore, we introduce lastFM_A1_dup, which is the same as lastFM_A1, except the size of each table is doubled by replicating each tuple once. This can show the impact of a higher result redundancy. Figures 11 (and 12) show the times for generating the join result and storing it on a disk (and the in-memory running time). Figure 13 and Figure 14 show the storage cost and the loading time for the aforementioned queries.

The results show that the two sources of inefficiency do not affect GJ, but they affect competitors significantly in both in-memory and in-disk runs. Note, the effect of UIR on Umbra is also not significant (as it is a WOJA), but higher result redundancy affects Umbra's performance. In terms of space, GJ is as expected drastically better. Interestingly, the loading time of GJ on lastFM_A1 is slower than MonetDB and Umbra. This is not the case for lastFM_A1_dup (larger redundancy) or for lastFM_A2 (higher UIRs). Finally, GJ performs better as join sizes increase, as Figures 11, 12, 13, and 14 show. The join sizes for lastFM_A1, lastFM_A1_dup and lastFM_A2 are ~61 million, ~493 million, and ~2 billion.

# 5 RELATED WORK

**Join Algorithms.** RDBMS products employ the well-known physical join algorithms (nested-loop, sort-merge and hash-join) and their derivatives [12–14, 20, 25, 30]). However, their performance has been found to be sub-optimal in certain cases [5, 53]. A lot of work has therefore been devoted to address these inefficiencies. [34–36] introduced join algorithms eliminating UIR, whose worst-case running time is proportional to the fractional edge cover. Leapfrog TrieJoin (LT), an influential WOJA, was introduced in [49]. Variations of LT for various other applications have been more recently developed, for example for SPARQL queries[22] and for commercial deductive (non-relational) DBMS (i.e., LogicBlox). Recent WOJAs also include [4] and [16]. More recently a new WOJA algorithm was developed [17] and introduced it within the Umbra system [32] – actually a new efficient implementation based on new data structures. [17] showed it is the state-of-the-art WOJA for RDBMS.

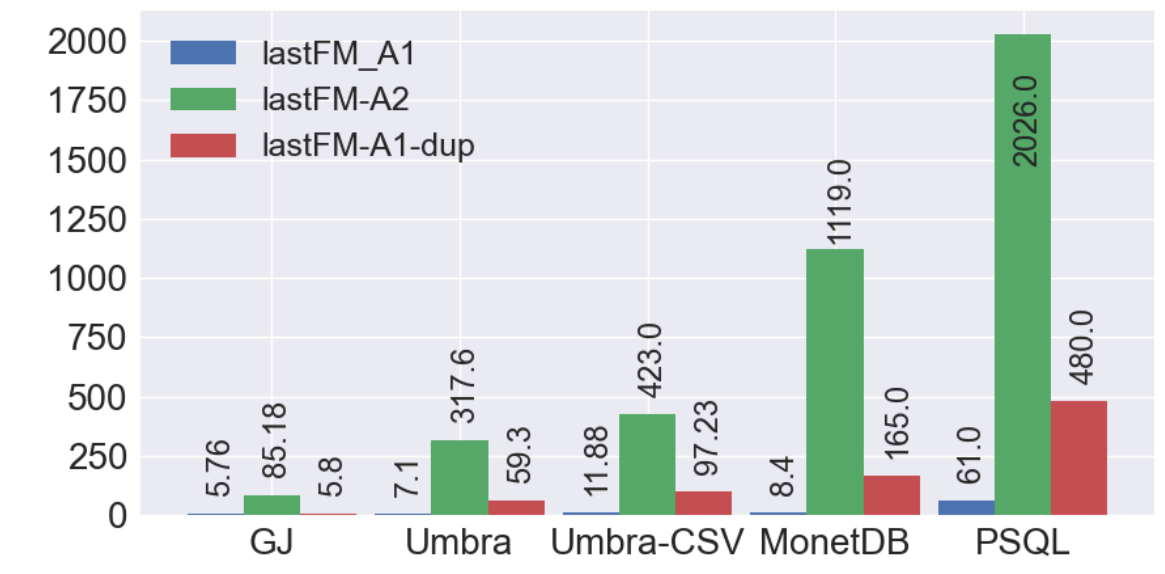

**Figure 11: Seconds for generating and storing the result**

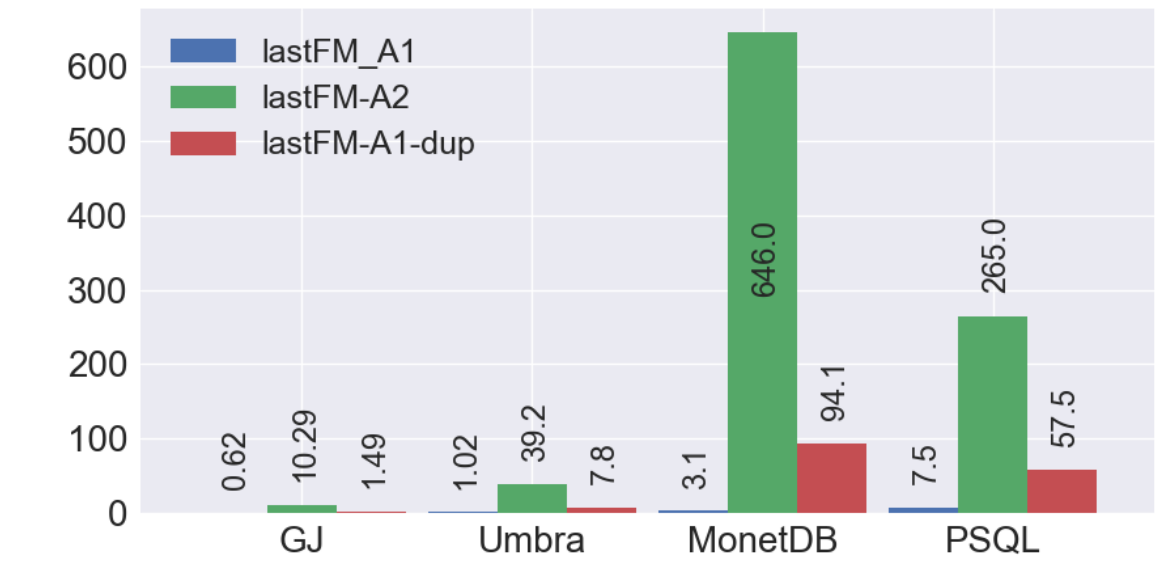

**Figure 12: Seconds for generating the result in memory**

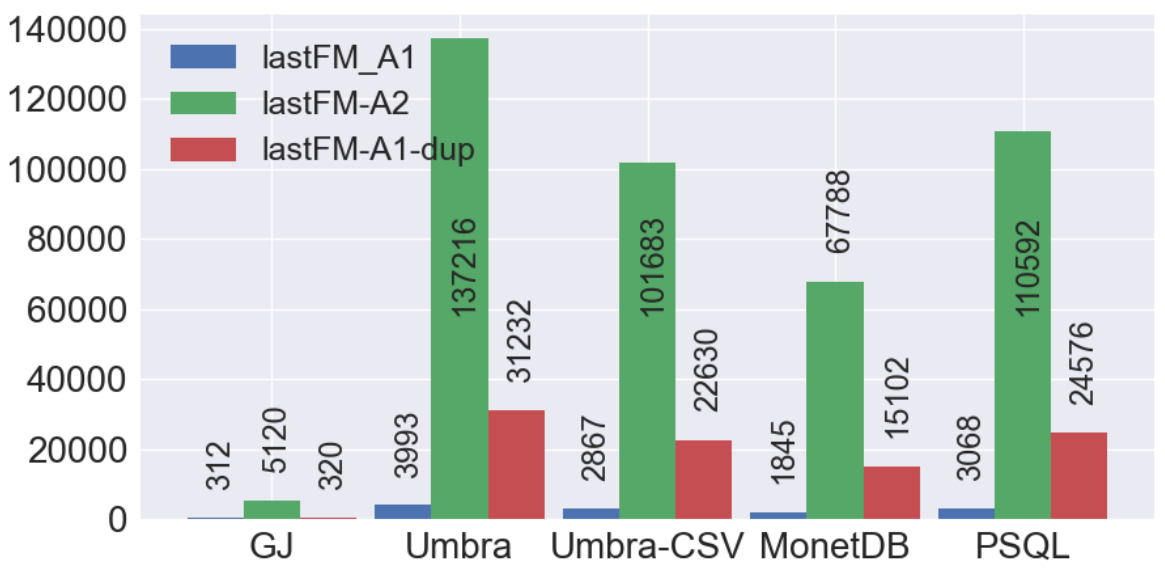

**Figure 13: Storage cost in MBs**

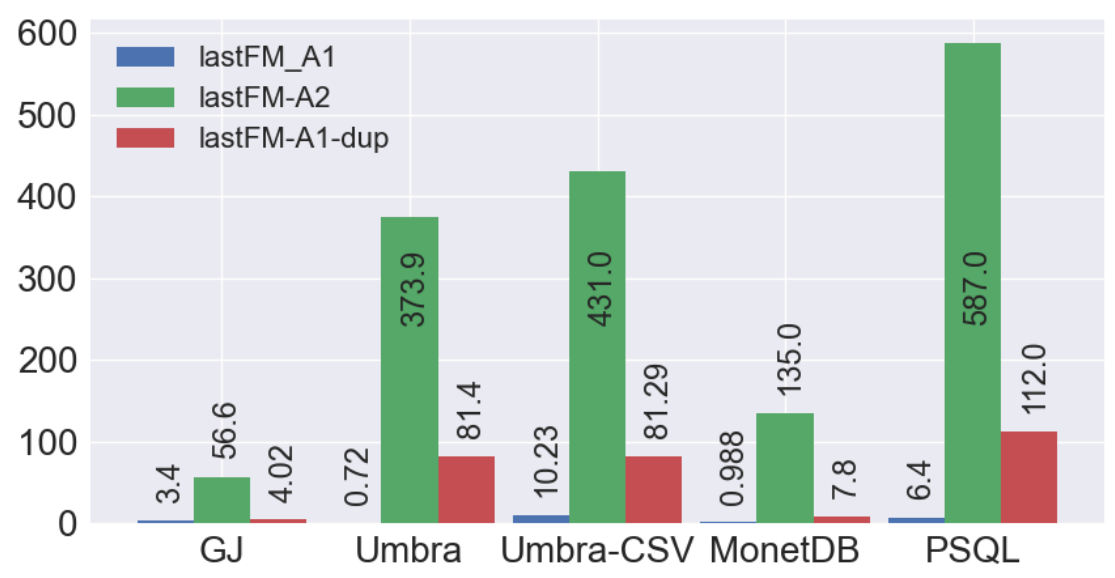

**Figure 14: Times in seconds for loading the results**

**Analytics over Join Results.** Another significant body of work focused on making analytics over join results more efficient, especially for large, expensive joins. This research thread is not concerned with producing more efficient join algorithms. There works fall into two major categories. First, works like [2, 9, 31, 37, 43, 44, 54]) aim to generate a uniform (or weighted) random sample of the join result *without having to generate the full join result.* These works are useful for approximate aggregation query results, or for training models over joins for purposes such as approximate query processing [28, 29], or for cardinality estimation [52]. PGMJoins [43] is the most recent among these and is similar to this work in that it uses PGMs. But PGMJoins is not (and was not meant to be) a physical join algorithm and lacks the inference algorithms needed to generate the full join result, factorize the join result and generate GFJS.

The second category centers on factorized DBs (FDBs) [33, 38–42]. These works are relevant to this work in that they can also offer a (albeit different type of) factorization of the join result. Their emphasis is on creating a high-level and un-materialized factorized join result to expedite downstream analytics tasks over join results, such as aggregations or regressions by avoiding generating the join result. We note that [39] does outline an enumeration algorithm for the join result tuples, showing that it is a "constant-delay" enumeration algorithm. Beyond this, no concrete, practical join algorithm is provided, as this approach was not viewed or proposed as a competitive join algorithm. It was not analyzed with respect to various types of queries and/or datasets and not compared against WOJAs or binary join plans.

It is also worth noting that GJ with its PGM-inspired approach leveraging potentials derives factorized `distributions of data`. In contrast, FDB derives a factorization of `data`. This is expected to have strong performance payoffs.

Finally, PGMs have also been used for computing aggregations over joins in [7] and [1], also offering different bespoke, tweaked inference algorithms. These works/algorithms are not and cannot be a physical join algorithm.

**Summarization of Join Results.** Different types of result redundancies are discussed in [18]. Summarizations for reducing redundancies are discussed in [3]. Various summary types are discussed in [46]. Even among these efforts, the proposed work is the first to generate a join summary (GFJS) without needing to first generate the join result.

Overall, to the best of our knowledge, this work is the first to derive a fresh approach for join query processing, putting forth its principled, PGM-inspired, summarization-desummarization approach, studying its performance in detail, comparing it against the state of the art join processing approaches, and showing that (and when) it is a preferable approach.

## 6 CONCLUSIONS

This work proposes and studies a drastically new approach, GJ, for n-way physical equi-joins in relational databases. GJ effectively tackles the fundamental inefficiencies of join algorithms and it is shown to be worst-case optimal. GJ advocates the generation of an RLE-style, frequency-based summary of the join result. It leverages PGMs and offers the inference algorithms needed to achieve this without computing the join first. GJ then proposes a new approach that consists of first computing the above summary, optionally storing it to disk (to be retrieved later when needed), and followed by desummarization to materialize the full result. Detailed experiments reveal that such an approach can yield large performance improvements compared against both traditional RDBMS (binary) join plans and WOJAs. Specifically, experiments with JOB, TPCH and lastFM data and queries showed GJ to be faster than PSQL up to 64X, more than 388X faster than MonetDB, and up to 6X faster than Umbra. GJ shines when large join results are to be stored on disk and reused (up to 820X faster than PSQL, up to 717X faster than MonetDB, up to 165X faster than Umbra and 94X faster than Umbra-CSV). GJ uses less storage than all competitors, up to 21,488X better than PSQL, up to 38,333X better than MonetDB, up to 78,750X better than Umbra and 54,666X better than Umbra-CSV. Because of this, GJ can actually produce the join result even in cases of very large join results where currently RDBMS join plans and WOJAs fail. Therefore, GJ can become a contender for join optimizers within RDBMSs, especially for joins yielding high UIR and large result redundancy.

## 7 ACKNOWLEDGEMENT

This work was supported by the UK Engineering and Physical Sciences Research Council (EPSRC) grant EP/R513374/1 for the University of Warwick